# Two-dimensional magnetotransport in $Bi_2Te_2Se$ nanoplatelets

*Pascal Gehring,[1a)] Bo Gao,[1a)] Marko Burghard,[1] and Klaus Kern[1,2]*

a) P.Gehring@fkf.mpg.de, b.gao@fkf.mpg.de

[1]Max Planck Institute for Solid State Research, Heisenbergstrasse 1, D-70569 Stuttgart, Germany

[2]Institute de Physique de la Matière Condensée, Ecole Polytechnique de Lausanne, 1015 Lausanne, Switzerland

Abstract

Single-crystalline $Bi_2Te_2Se$ nanoplates with thicknesses between 8 and 30 nm and lateral sizes of several micrometers were synthesized by a vapour-solid growth method. Angle-dependent magnetoconductance measurements on individual nanoplates revealed the presence of a two-dimensional weak anti-localization effect. In conjunction with gate-dependent charge transport studies performed at different temperatures, evidence was gained that this effect originates from the topologically protected surface states of the nanoplates.

Three-dimensional topological insulators (TIs) are bulk insulators but posses topologically protected Dirac-like surface states.[1,2,3] The spin of the charge carriers in these states is locked to their momentum, rendering them immune against non-magnetic scattering centres.[4] Owing to this property, TIs are interesting for both, fundamental research and spintronic applications. However, while a range of compounds have been theoretically predicted[5,6,7,8,9,10] and experimentally proven by angle-resolved photoemission spectroscopy (ARPES)[1,2,3,11,12,13] to be TIs, such proof turned out to be more elusive by electrical transport measurements. This situation mainly arises from the high conductivity of the intrinsically doped bulk, making it difficult to separate surface and volume effects.[2,14,15,16] One approach to overcome this problem is to counter-dope the material and thus shift the Fermi level into the bulk bandgap.[4,17,18,19] Further options are to increase the surface-to-volume ratio,[20,21] to apply electrical gating in order to reduce the carrier density,[21,22] or to enhance the surface transport contribution by using materials with low bulk mobility. With respect to the latter possibility, one of the most promising candidates is $Bi_2Te_2Se$ that has been theoretically predicted[23] and experimentally confirmed by ARPES[24] to be a TI. According to recent magnetotransport studies, $Bi_2Te_2Se$ has indeed a high bulk resistivity of 6 $\Omega$cm, and hence a strong surface contribution to the total conductance of about 6%, which is the largest value thus far reported for TIs.[25] In the present work, we investigate thin platelets of this compound, in particular the possibility to control the position of the Fermi level in the bulk band gap via electrostatic gating. Such capability is needed to compensate for the fact that most of the established synthesis procedures yield samples wherein the bulk dominates the electrical transport.

The $Bi_2Te_2Se$ nanoplates were synthesized by a catalyst-free vapour-solid (VS) method[21]. To this end, ultrapure $Bi_2Se_3$ and $Bi_2Te_3$ crystalline powders were placed in the hot zone of a horizontal tube furnace (tube diameter 2.5 cm), with $Si/SiO_x$ substrates being placed about 15 cm away within the colder downstream region. The tube was repeatedly evacuated to a pressure of $p < 1$ mbar and flushed with ultrapure argon. Subsequently, the carrier gas flow rate and pressure were adjusted to 150 sccm and 80 mbar, respectively. The furnace was then heated to 590°C, at which temperature it was kept for 6 min, followed by natural cool down without gas flow at a constant pressure of 80 mbar. It was found that the stoichiometry $Bi_2(Se_xTe_{1-x})_3$ of the product depends sensitively on the position of the $Bi_2Se_3$ and the $Bi_2Te_3$ sources and the molar ratio of the powders used. In order to obtain $Bi_2Te_2Se$, 266 mg of $Bi_2Se_3$ powder (hot zone) and 355 mg of $Bi_2Te_3$ powder (6cm away) had to be used. According to atomic force microscopy (AFM) and scanning electron microscopy (SEM) analysis, thus obtained $Bi_2Te_2Se$ nanoplates have lateral sizes of 1-10 μm and thicknesses of 8-30 nm.

In **Fig. 1a,** a transmission electron microscopy (TEM) image and a corresponding selected area diffraction pattern (inset) of a typical $Bi_2Te_2Se$ platelet are shown. $Bi_2(Se_xTe_{1-x})_3$ forms rhombohedral (space group $R\bar{3}m$) crystals that consist of hexagonally close-packed atomic layers of five atoms (quintuple layer) which arrange along the c-axis as follows: $Se^{(1)}/Te^{(1)} - Bi - Se^{(2)}/Te^{(2)} - Bi - Se^{(1)}/Te^{(1)}$. All flakes investigated by TEM were found to be single crystalline, exhibiting {0001} and {1120} crystal facets. Toward determining their chemical composition, the $Bi_2(Se_xTe_{1-x})_3$ platelets were investigated by Raman microscopy. The chalcogenides $Bi_2Se_3$ and $Bi_2Te_3$ are strongly Raman active and can be identified by their characteristic $A^1_{1g}$, $E_{2g}$ and $A^2_{1g}$ Raman peaks in the low wavenumber region.[26] In the corresponding alloys $Bi_2(Se_xTe_{1-x})_3$ these peaks shift with changing composition.[27] For x < 1/3 this shift is small, since only the $Te^{(2)}$ atoms are replaced by Se atoms and the $Bi-Te^{(2)}$ and $Bi-Se^{(2)}$ bonds are of similar strength. In contrast, when x exceeds 1/3 also $Te^{(1)}$ atoms get replaced by Se, which causes a notable shift of the $A^1_{1g}$ mode, as well as a splitting of the $A^2_{1g}$ mode due to out-of-phase movements of the outer Bi and $Te^{(1)}/Se^{(1)}$ atoms. **Fig. 1b** compares the Raman spectra acquired from the $Bi_2Se_3/Bi_2Te_3$ starting materials and a representative $Bi_2(Se_xTe_{1-x})_3$ nanoplate. The experimentally accessible range of 80 - 200 cm$^{-1}$ contains the $E_{2g}$ and $A^2_{1g}$ peaks, but not the $A^1_{1g}$ peak. The nanoplate spectrum displays three distinct features, namely the $E_{2g}$ peak at 105.2 cm$^{-1}$ and the $A^2_{1g}$ peak which is split into two components at 138.8 and 148.9 cm$^{-1}$, respectively. The latter two peak positions are in good agreement with the corresponding values of 139.7±1 cm$^{-1}$ and 148.5±1 cm$^{-1}$, as reported by Richter *et al.*[27] for $Bi_2Te_2Se$ bulk samples, from which x = 0.34 can be concluded for the present samples.

In order to determine the charge transport mechanism and dimensionality of the involved transport channels in the nanoplatelets, we performed magnetotransport studies. To this end, the $Bi_2Te_2Se$ nanoplates were mechanically transferred onto Si substrates covered with a 300 nm thick $SiO_2$ layer. Individual plates were provided with Ti(4 nm)/Au(40 nm) contacts in Hall-bar or van-der-Pauw geometry using e-beam lithography (as exemplified in **Fig. 1c**). Prior to metal evaporation, the exposed regions were subjected to Ar-plasma (50 s at 250 W) in order to reduce the contact resistance. Low temperature Hall measurements on several devices revealed an average electron density $n_{3D}$ of about $10^{25}$ m$^{-3}$, a value comparable to typical Bi$_2$Se$_3$ crystals.[21] The average Hall mobility µ of our samples was determined to be on the order of 100 - 400 cm$^2$/Vs, lower than in pure Bi$_2$Se$_3$ or Bi$_2$Te$_3$ nanostructures[33,35]. **Fig. 2a** presents the low field magnetoconductance signal $\Delta\sigma = \sigma(B) - \sigma(B=0)$ of a 15 nm thick Bi$_2$Te$_2$Se plate at T = 1.5 K as a function of the tilting angle θ between the z-axis and the B-field direction. The prominent peak around B = 0 T can be attributed to the weak anti-localization (WAL) effect, which is characteristic of materials wherein strong spin-orbit coupling strongly suppresses backscattering due to time-reversal symmetry, resulting in a negative correction to the resistance at zero B-field[33]. Application of a magnetic field breaks the time reversal symmetry, leading to enhanced backscattering and a corresponding increase of resistance. This effect is not limited to 2D systems, but can also have a contribution from the 3D bulk. To test whether the observed WAL effect is a pure two-dimensional (2D) effect the sample was tilted in the external magnetic field, since in this case the signal should depend only on the magnetic field component $B_z$ = Bsin(θ) normal to the sample surface. For the present samples, the WAL effect vanishes at θ = 0° (B in plane, $B_z$ = 0), under which condition the magnetoconductance shows a simple parabolic behavior like in conventional semiconductors. The 2D character of the WAL is further illustrated by **Fig. 2b**, where $\Delta\sigma$ is plotted as a function of $B_z$. It can be seen that all curves coincide for different angles θ. The observed angular dependence of $\Delta\sigma$ signifies the 2D character of the WAL, suggesting that this effect originates from the topologically protected 2D surface states. A similar conclusion based upon angle-dependent magnetotransport measurements has recently been drawn for thin Bi$_2$Te$_3$ films grown by MBE.[28] The magnetoconductance data can be well fitted by the Hikami-Larkin-Nagaoka (HLN) model[29] for 2D localization:

$\Delta\sigma(B) = \alpha(e^2/h)[\ln(B_\phi/B) - \Psi(1/2 + B_\phi/B)]$, with $B_\phi = \hbar/(4el_\phi^2)$ where $\Psi$ is the digamma function, $l_\phi$ is the phase coherence length, and the constant $\alpha = 0.5$ for each contributing 2D transport channel. By fitting the magnetoconductance curve at θ = 90° with the HLN equation yields $\alpha = 0.57$ and $l_\phi = 69.5$ nm. The value of α = 0.57 ≈ 0.5 indicates that only one surface is contributing to the 2D WAL effect, in analogy to $Bi_2Se_3$[30] and $Bi_2Te_3$[28] thin films. It is noteworthy, that the Δσ(B) signal in **Fig. 2a** shows a notable oscillatory behavior. These features could be explained by quantum interference phenomena like universal conductance fluctuations[36] or the Aharonov-Bohm effect[37] which have been observed in various TI materials.

Having demonstrated the pure 2D magnetotransport in very thin nanoplatelets, we explored the possibility of using an external gate to tune the Fermi level in such manner as to favor charge transport through the surface state also in thicker platelets. Such samples still show WAL at zero angle,[28] and their Fermi level position is expected near the conduction band edge. Along these lines, we measured the temperature dependence of resistance at different back gate voltages for several platelets with a thickness above 20 nm. The aim of these experiments was to access the Fermi level position from the observed thermal activation barriers. The behavior is exemplified in **Fig. 3a** for a ~20 nm thick $Bi_2Te_2Se$ nanoplatelet. While at zero gate voltage, purely thermally activated behaviour can be discerned, at highest negative gate voltage (-110 V) a resistance maximum emerges at approximately 50 K. At moderate temperatures, the mobility of the surface electrons is expected to be low due to strong electron-phonon scattering,[31] and correspondingly the charge transport through the bulk should dominate over the surface state. On this basis, the thermal activation behaviour can be attributed to the excitation of surface state electrons nearby the Fermi level into the bulk conduction band (CB), with the excitation energy being equal to the energy gap Δ between the Fermi level and the CB edge.[32] We extracted the energy gap Δ by fitting the R(T) curves in the high T range (80 - 180 K) with an Arrhenius formula $R(T) = R_0 + A*\exp[-\Delta/T]$. (black, dotted curves in **Fig. 3a**). Thus obtained values are plotted in **Fig. 3b** as a function of gate voltage. It is apparent that the application of increasingly negative gate voltages results in the expected increase of the energy gap. The maximum in the R(T) curves at negative gate voltages can be explained by an interplay between two effects. The first one comprises a resistance increase due to the reduced carrier concentration at lower temperatures. The second effect, which counteracts the aforementioned resistance increase, involves the reduction of electron-phonon scattering of the surface electrons upon cooling, owing to the metallic character of surface state electrons.[32] As distinguished from the behaviour under negative gate voltages, the temperature behaviour can be changed into purely metallic by applying a strong positive gate voltage (+80 V). In this gating regime, the Fermi level is located within the bulk conduction band, similar to quasi-metallic $Bi_2Se_3$ crystals without compensation doping[34].

In summary, we have demonstrated the presence of a 2D WAL effect in thin nanoplatelets of $Bi_2Te_2Se$, which points toward the participation of the topologically protected surface states in the charge transport. The samples are amenable to gate control, as proven by gate- and temperature-dependent resistance measurements, enabling the manifestation of surface state transport also for thicker platelets. Our results establish $Bi_2Te_2Se$ as a valuable material for further studies of the fundamental properties of topological surface states.

**Acknowledgements**

We wish to acknowledge G. Pathak for experimental support, as well as Y. Weng for help with the TEM experiments.

**Figure captions**:

**Figure 1:** a) Transmission electron microscopy (TEM) image and corresponding selected area diffraction pattern (inset) of a $Bi_2Te_2Se$ nanoplate. Scale bar corresponds to 1 μm (inset: 5 1/nm) b) Raman spectra of a $Bi_2Se_3$ flake (top), a $Bi_2Te_2Se$ nanoplate synthesized by the CVD method (middle), and a $Bi_2Te_3$ flake (bottom). c) Optical image of a 15 nm thick flake contacted in Hall bar geometry (scale bar: 5 μm).

**Figure 2:** Corrected low field magnetoconductance signal acquired from a 15 nm thick $Bi_2Te_2Se$ nanoplate for different angles θ as a function of (a) the magnetic field and (b) the magnetic field component $B_z = B\sin(\theta)$ normal to the surface. The fitting result for the WAL curve recorded at θ=90° is shown in black. For θ=0° the data was fitted with a $B^2$ model.

**Figure 3:** a) Temperature dependent resistance recorded at different backgate voltages applied to a $Bi_2Te_2Se$ nanoplate with a thickness of ~ 20 nm. Dotted lines: Arrhenius fits to the curves. b) Energy gap Δ between Fermi level and bulk conduction band edge for different gate voltages, as obtained from Arrhenius type fitting. The inset shows the definition of Δ.

**Figure 1**

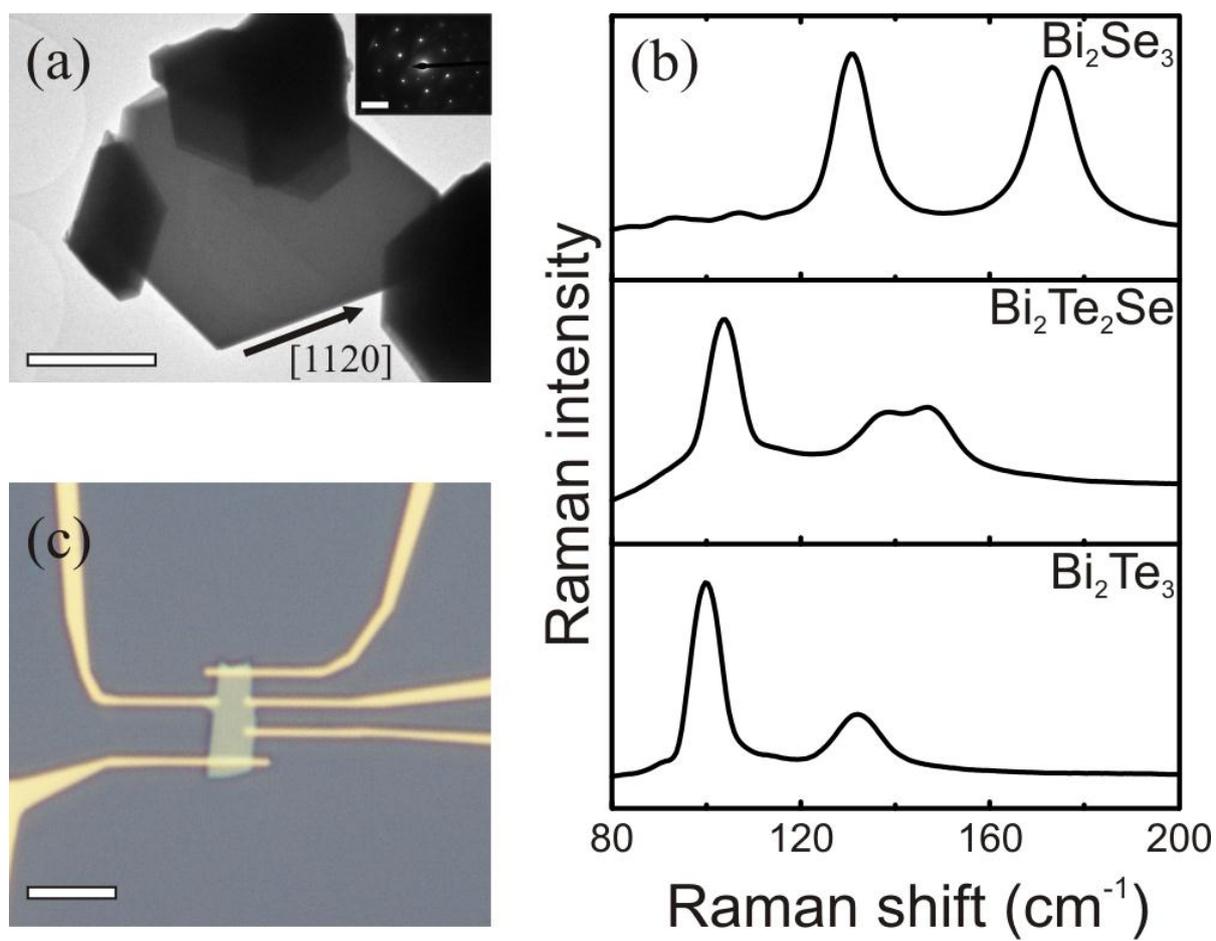

**Figure 2**

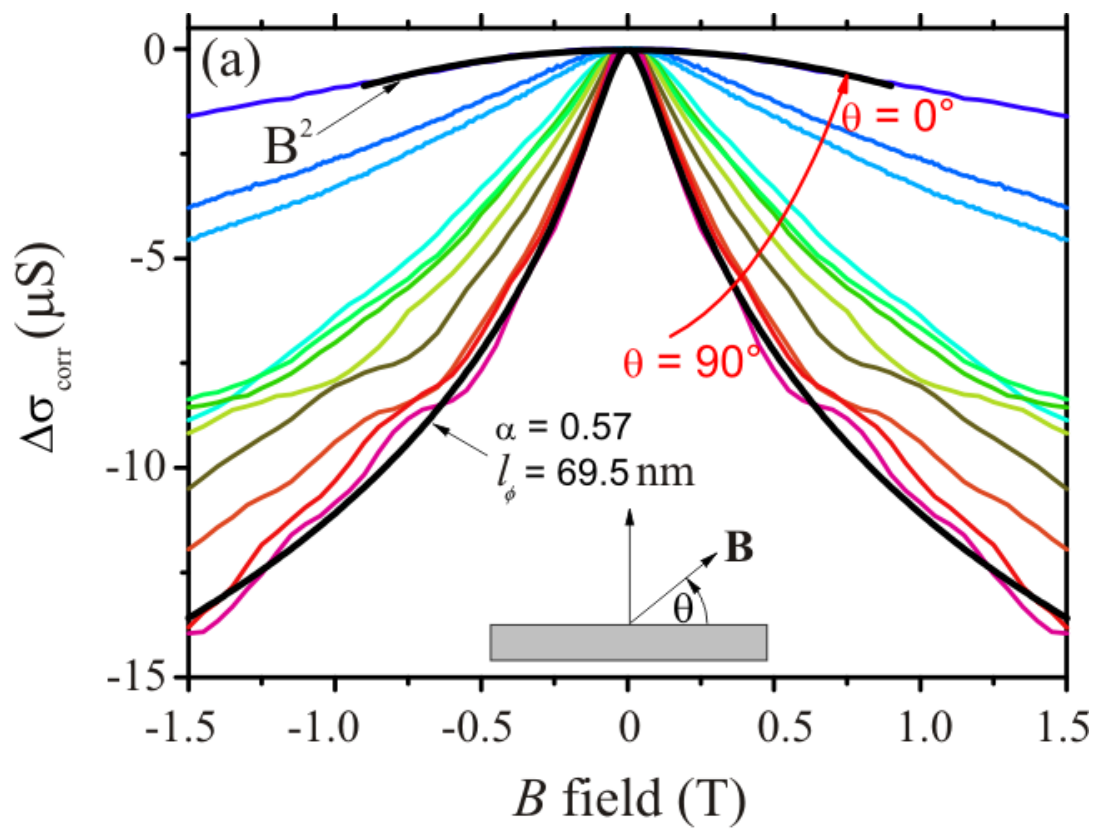

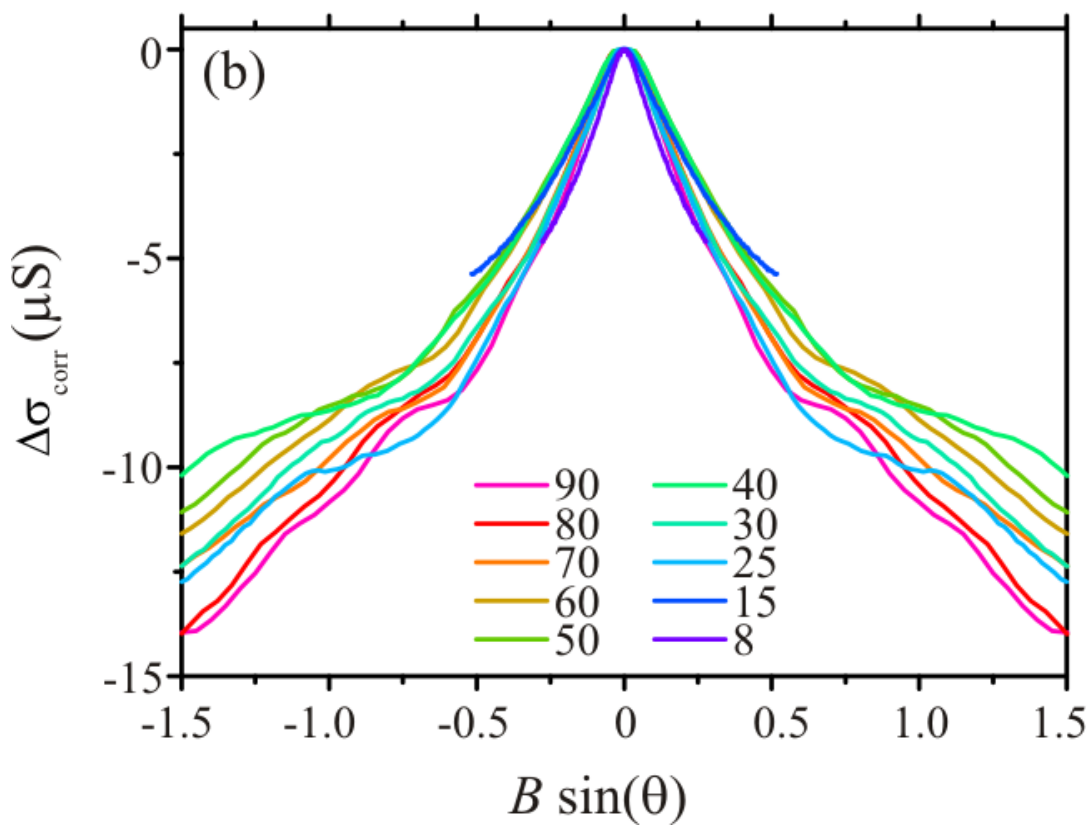

**Figure 3**

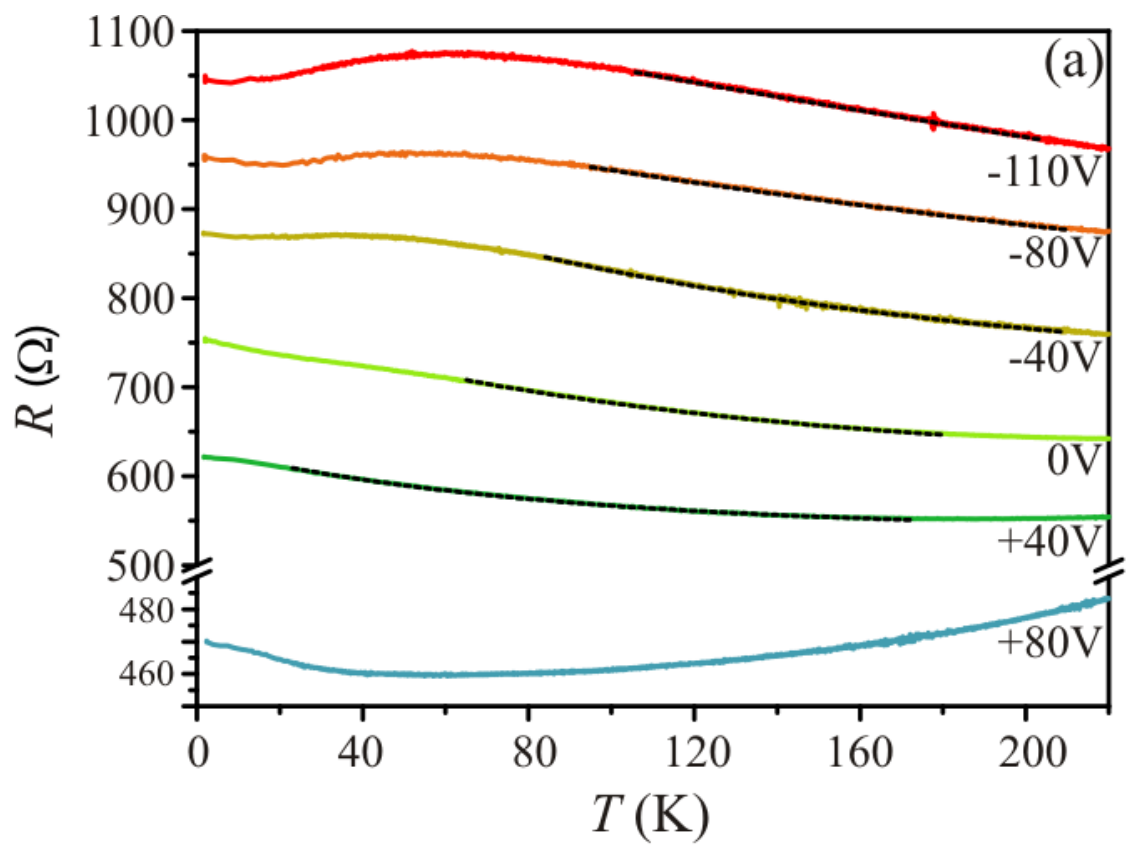

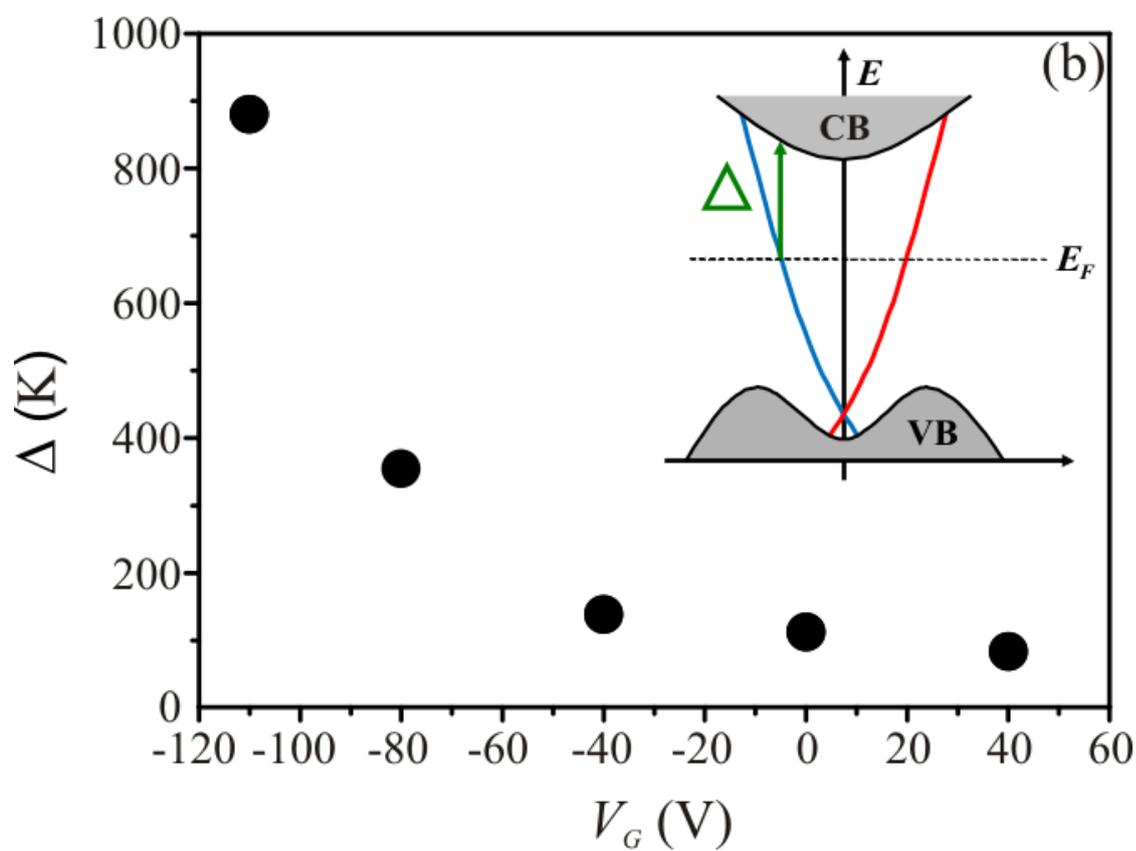